\documentclass[aps,prd,twocolumn,showpacs,nofootinbib,preprintnumbers]{revtex4-1}

\usepackage{amsmath,amssymb,bm}
\usepackage[dvips]{graphicx}
\usepackage{hyperref}
\usepackage{yfonts}
\newcommand{\beq}{\begin{eqnarray}}
\newcommand{\eeq}{\end{eqnarray}}
\newcommand{\SU}{\text{SU}}

\begin{document}

\title{New critical point for QCD in a magnetic field}

\author{Thomas D. Cohen}
\author{Naoki Yamamoto}
\affiliation{Maryland Center for Fundamental Physics, 
Department of Physics, University of Maryland,
College Park, Maryland 20742-4111, USA}

\begin{abstract}
We provide a general argument for the possible existence of a new critical point associated 
with a deconfinement phase transition in QCD at finite temperature $T$ and in a 
magnetic field $B$ with zero chemical potential. 
This is the first example of a QCD critical point in a physical external parameter region
that can be studied using lattice QCD simulations without suffering from a sign problem.
\end{abstract}
\maketitle

Much attention has been focused on the properties of QCD at high temperature 
$T$ and/or baryon chemical potential $\mu_B$ to understand various phenomena 
including heavy ion collisions, neutron stars, and the early Universe. 
Such systems are often subject to the effects of a strong magnetic field 
$B$. Indeed, the strong magnetic field of order the QCD scale 
$\Lambda_{\rm QCD}$ (or $10^{18}$--$10^{19}$ Gauss) may be reachable in 
noncentral heavy ion collision experiments at RHIC and LHC \cite{Kharzeev:2007jp} 
and possibly \emph{inside} magnetars \cite{Magnetar}.
Moreover, the strong magnetic field may have existed in the early Universe as an 
origin of the present large-scale cosmic magnetic field \cite{Grasso:2000wj}.
It is thus important to unravel the possible modifications of the QCD phase 
diagram in the presence of a strong magnetic field. 

Unlike QCD with nonzero $\mu_B$,  QCD with nonzero $B$  does not have a sign 
problem.  Thus the QCD phase diagram in the $(T, B)$ plane can be determined 
from first principles from Monte Carlo calculations of lattice QCD using existing 
techniques. For the chiral transition, recent lattice QCD results \cite{Bali:2011qj} 
indicate a discrepancy with earlier model results \cite{Mizher:2010zb} 
for the qualitative behavior of the critical temperature as a function of $B$, raising 
a question on the reliability of the conventional model analyses. A possible 
explanation of this discrepancy was proposed in Refs.~\cite{Fukushima:2012kc}.

\begin{figure}[t]
\begin{center}
\includegraphics[width=6cm]{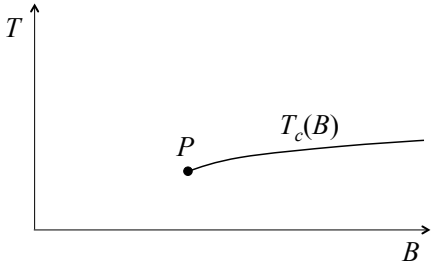}
\end{center}
\vspace{-0.5cm}
\caption{Putative phase diagram in the $(T, B)$ plane. $T_c(B)$ denotes the critical 
temperature of the first-order deconfinement transition as a function of $B$ that 
ends at the critical point $P$. This figure is to be understood only topologically.}
\label{fig:phase}
\end{figure}

This paper concerns the confinement/deconfinement transition 
in the $(T, B)$ plane at zero chemical potential. Naively, one might think that, as the 
magnetic field does not couple to gluons directly, its effect on confinement will  not 
be dramatic and might be very difficult to understand in a theoretically controlled manner 
(see Refs.~\cite{Galilo:2011nh, Fraga:2012ev, Anber:2013tra} for recent attempts). 
However the phase structure associated with the physics of
confinement in the presence of the magnetic field is qualitatively novel, and, given a 
single plausible assumption, may be understood in a controlled way. 
Our main result of the phase diagram on the $(T, B)$ plane is summarized in Fig.~\ref{fig:phase}. 
In particular, we argue that it is highly plausible that a new critical point for the 
\emph{deconfinement} phase transition (denoted by $P$ in Fig.~\ref{fig:phase}) exists.  
This argument does not depend strongly on model-dependent analysis.%
\footnote{A possible magnetic critical point was also suggested in a model calculation
\cite{Agasian:2008tb}. However, in their phase diagram, the first-order 
deconfinement transition is turned into a crossover with increasing $B$,
which is different from ours.} 
To distinguish it from a possible conventional QCD critical point at finite 
$\mu_B$ \cite{Stephanov:2004wx} (see also Ref.~\cite{Hatsuda:2006ps})---which is thought 
to be associated with approximate chiral symmetry---we shall call this the ``magnetic critical point."   
For a possible conventional critical point in a magnetic field, see, e.g., Refs.~\cite{Andersen:2012bq}.

We note that the critical point under discussion has a fundamental virtue compared to 
the usual QCD critical point associated with chemical potential and temperature.  
The conventional critical point, if it exists, is a property of QCD itself. 
However, when electromagnetic effects are included the thermodynamics are fundamentally 
altered: one cannot have an infinitely extended charged phase due to energetics.  
In contrast the critical point here includes electromagnetic effects---indeed, it depends on them.

Throughout the paper, we assume the magnetic field to be homogeneous and in the $\hat{z}$ 
direction with magnitude $B$. For notational simplicity, we first put the current quark mass 
$m_q$ to zero, but generalization to nonzero $m_q$ is trivial, as we shall mention later.
We also put chemical potential $\mu_B$ to zero unless otherwise stated.

The argument for the existence of the first-order transition with $T$ at large $B$ is 
quite straightforward.  The phenomenon of magnetic catalysis of chiral symmetry breaking, 
the strengthening of the chiral condensate in an external magnetic field, is well known
at $T \ll \sqrt{eB}$ \cite{Shovkovy:2012zn}.%
\footnote{It should be remarked that the inverse magnetic catalysis is observed for 
$T \sim \sqrt{eB} \sim \Lambda_{\rm QCD}$ in lattice QCD simulations \cite{Bali:2011qj}. 
In the regime $T \ll \sqrt{eB}$, however, \emph{both} lattice QCD simulations 
at $\sqrt{eB} \sim \Lambda_{\rm QCD}$ \cite{Bali:2011qj} and the controlled 
weak-coupling analysis at $\sqrt{eB} \gg \Lambda_{\rm QCD}$ \cite{Miransky:2002rp}
\emph{do} observe the magnetic catalysis. 
In this paper, we shall only make use of the magnetic catalysis in the latter regime
(which is not reachable in the present lattice calculations).}
Accompanying this phenomenon is an increase of the effective mass of quarks \cite{Miransky:2002rp}.  
In the regime of very large magnetic field, the quarks become sufficiently heavy 
that they decouple from the gluodynamics, yielding an effective theory of pure 
gluodynamics---albeit of an anisotropic type.  Thus, center symmetry is an emergent 
symmetry of QCD at large magnetic fields, since the effective theory is invariant under 
center transformations.    

Consider this lowest-order effective theory as a theory in its own right and assume that 
center symmetry is unbroken as $T \rightarrow 0$ in this theory.  It is known to be 
broken at high temperatures.  Thus, at the level of the lowest-order effective theory, 
there must be a phase transition.  Moreover, there is reason to believe that this 
transition is first order. Note that there is a general argument that conventional
(i.e. isotropic) gluodynamics should not have a second-order phase transition 
\cite{Yaffe:1982qf}; lattice simulations confirm that pure gluodynamics does indeed 
have a first-order transition for the isotropic case \cite{Fukugita:1989yb}.

Since the lowest-order effective theory is a pure glue theory, it is natural to expect that it has 
a first-order transition---if for no other reason than pure glue theory has a first-order 
transition in the isotropic case.%
\footnote{One might be concerned that the pure gluodynamics in the large $B$ limit acts like 
a dimensionally reduced theory which shows a second-order transition \cite{Yaffe:1982qf}. 
However, the theory in this regime is not in (2+1) dimensions, but in (3+1) dimensions with 
\emph{anisotropic} color dielectric constant but with \emph{isotropic} color magnetic 
permeability [see Eq.~(\ref{leff})]. It is thus important to check our assumption of 
first-order transition for this anisotropic theory on the lattice (see also below).} 
Moreover, first-order transitions are, by their nature, robust: small changes in the details 
of a theory cannot destroy the transition due to the existence of a nonzero latent heat.
In this respect first-order transitions are quite different from second-order ones.
An arbitrarily small change in the details of the theory can completely eliminate a 
second-order transition by turning it into a crossover.
Since the effective theory has a first-order transition, and at sufficiently large $B$ is 
equivalent to the lowest-order effective theory up to small corrections, QCD too 
will have a first order transition. Given the existence of a first-order transition at large 
$B$ and only crossover behavior at $B=0$, there must be some minimum value of $B$ 
for which the transition occurs. The most natural way for this to occur is simply via 
a critical point as in Fig.~\ref{fig:phase}.

The simple argument given above depends on the behavior of  QCD in the regime 
$eB \gg \Lambda_{\rm QCD}^2$.   Here we briefly recapitulate the known physics 
in this regime using the analysis of Ref.~\cite{Miransky:2002rp}.
In this regime, the quark dynamics at low energy is dominated by the lowest Landau level (LLL).  
Also the QCD coupling constant, $\alpha_s$, can be shown to be sufficiently small 
enough that the calculations are under theoretical control. 
A self-consistent gap equation for the quarks in the LLL---similar to 
(color) superconductivity or superfluidity at large chemical potential \cite{Son:1998uk}---can
be derived and solved, yielding a quark mass gap $M_{\rm dyn}$ \cite{Miransky:2002rp},
\beq
\label{mass-gap}
M_{\rm dyn} = C(\alpha_s) \sqrt{|e_qB|},
\eeq
where $e_q$ are the charges of the quarks, 
$(e_u, e_d, \cdots)=\left(\frac{2}{3}, -\frac{1}{3}, \cdots \right)e$. 
The detailed expression for $C(\alpha_s)$ based on the consistent truncation of 
the gap equation \cite{Miransky:2002rp} is irrelevant for our purpose; what will 
be important for us is that $M_{\rm dyn}$ is an increasing function of $B$ at 
sufficiently large $B$ and $M_{\rm dyn} \rightarrow \infty$ for $B \rightarrow \infty$. 
Note that $\alpha_s \ll 1$ is indeed valid for $M_{\rm dyn}^2 \ll k^2 \ll eB$, 
where $k$ is the typical momentum in the gap equation.

That quarks acquire a large mass gap in a strong magnetic field means that 
quarks decouple from the gluons at low energy scales---well below $\sqrt{eB}$.  
Hence, the lowest-order effective theory for low-energy dynamics is described by an 
anisotropic pure  $\SU(3)$ gauge theory with an effective Lagrangian of the form
\begin{align}
&{\cal L}_{\rm eff}^0 = -\frac{1}{4} F^a_{\mu \nu} \Gamma^{\mu \nu}_{\, \, \alpha \beta} 
F^{a \, \alpha \beta}  \; \; \; {\rm with} \label{leff} 
\\ 
&\Gamma^{\mu \nu}_{\, \, \alpha \beta} \equiv 
g^\mu_{\, \alpha} g^\nu_{\, \beta} + (\epsilon_{z z}-1) (\delta^{\mu 3} \delta^{\nu 0}\delta_{\alpha3} \delta_{\beta 0} +\delta^{\mu 0} \delta^{\nu 3}\delta_{\alpha 0} \delta_{\beta 3}), \nonumber
\end{align}
where $\epsilon_{zz}$ is known to be much larger than unity \cite{Miransky:2002rp}; 
it is the only term in the dielectric tensor which differs from its vacuum value.   
The anisotropy is induced by the magnetic field which breaks rotational invariance.  
The superscript $0$ on ${\cal L}_{\rm eff}$ is to indicate that this is the lowest term in 
an expansion for the effective theory.  Higher-order terms will be suppressed by factors 
of $p/M_{\rm dyn}$  where $p$ is a characteristic momentum being probed; these terms 
become negligibly small at large $B$.

Note that the effective theory in Eq.~(\ref{leff}) contains covariant derivatives in the field 
strengths and through these, the coupling constant enters.  As in other non-Abelian 
gauge theory the coupling constant is scale dependent and the effective 
theory acquires a scale through dimensional transmutation.  Its scale differs from the 
conventional QCD scale, $\Lambda_{\rm QCD}$. This is because the \emph{effective} 
coupling constant $\alpha_s'$ of this effective theory is different from 
that of the original QCD, $\alpha_s$, and it is defined such that 
$\alpha_s'(M_{\rm dyn}) = \alpha_s(\sqrt{eB})$; the resultant scale 
$\Lambda_{\rm QCD}'(B)$ is much smaller than $M_{\rm dyn}(B)$ \cite{Miransky:2002rp}. 
A similar reduction of the confinement scale has also been argued for a two-flavor color 
superconductor \cite{Rischke:2000cn}.

As noted above, the existence of this pure glue effective theory ensures that the 
leading-order effective theory must have a phase transition, which may be naturally taken
first order.  Such a first-order transition implies that QCD at sufficiently large $B$ also 
has a first-order transition. The phase transition temperature in this regime will be fixed 
by the scale of the effective theory, $\Lambda_{\rm QCD}'(B)$.  
On the other hand, at $B=0$, it has been established from lattice QCD studies that 
there is no  first-order phase transition \cite{Aoki:2006we}; the deconfinement regime 
emerges as a result of a crossover. 
Therefore, the line of first-order deconfinement transitions at large $B$ above has to 
terminate at some point at some critical value of the magnetic field which we denote 
$B_c$. The most natural way for this to occur is for it to terminate at a critical point 
in the $T$-$B$ plane---$(T_c,B_c)$ as in Fig.~\ref{fig:phase}.  
Such a critical point of a line of first order is a point at which a second-order transition 
takes place.  

Although we cannot infer its location from our argument alone, 
we can estimate its scale: $\sqrt{eB_c} \sim T_c \sim \Lambda_{\rm QCD}$ 
in the chiral limit, as $\Lambda_{\rm QCD}$ is the only relevant scale.
Note that the scale $\Lambda_{\rm QCD}$ does not necessarily mean 
it is just around 200 MeV; it can be larger than 1 GeV, see the discussion below.

\begin{figure}[t]
\begin{center}
\includegraphics[width=4.5cm]{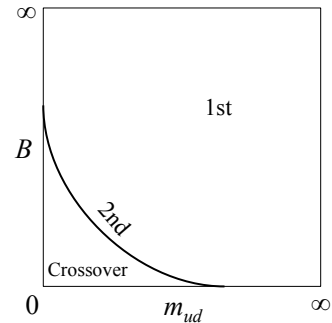}
\end{center}
\vspace{-0.5cm}
\caption{Finite-$T$ deconfinement transition in the $(m_{ud}, B)$ plane.}
\label{fig:columbia}
\end{figure}

If we turn on nonzero quark masses, quarks decouple earlier from gluodynamics 
with increasing $B$; the phase diagram in the $(m_{ud}, B)$ plane, similar to the 
Columbia plot, is shown in Fig.~\ref{fig:columbia} 
(for two degenerate massive flavors, as an example).

We note that this analysis depends on our assumption that the leading-order 
effective theory at large $B$ has a first-order transition. As remarked above, it is 
important to verify this assumption directly via lattice studies of the (anisotropic) 
leading-order effective theory in Eq.~(\ref{leff}). Such studies are relatively straightforward to 
conduct since they do not involve fermions and hence do not require the calculation of a 
determinant. We also note that this analysis does \emph{not} rely on the magnetic catalysis 
at $\sqrt{eB} \sim \Lambda_{\rm QCD}$, and whether the inverse magnetic catalysis 
occurs there is irrelevant. What is used is that the effective quark mass increases 
at $\sqrt{eB} \gg \Lambda_{\rm QCD}$ (and $T \ll \sqrt{eB}$).

Given the assumption that the leading-order effective theory has a first-order 
transition at large $B$, the existence of a critical value $B_c$ below which 
the transition vanishes is assured.  However, as a logical matter, there is no 
guarantee that the phase diagram must be of the form of  Fig.~\ref{fig:phase}.  
Other possible scenarios exist.  

\begin{figure}[t]
\begin{center}
\includegraphics[width=6cm]{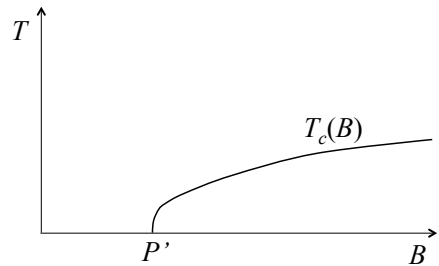}
\end{center}
\vspace{-0.5cm}
\caption{A possible phase diagram where the 
first-order deconfinement transition terminates at $P'$. }
\label{fig:phase2}
\end{figure}

One possible class of alternative scenarios involves the possibility that phase diagram is 
more complicated so that for some values of $B$ more than two phases exist. 
We believe that such scenarios are unlikely to be correct. In any event, such 
scenarios share with the simple one in Fig.~\ref{fig:phase} the feature that 
at $B_c$ there is a critical point. Thus, the existence of a critical point seems quite robust. 

However, there is one class of scenarios that cannot be excluded in which a critical point 
of this sort does not occur. In these scenarios, rather than the first-order line ending in a 
second-order critical point, it remains first order all the way to the end and terminates 
at $T=0$ (Fig.~\ref{fig:phase2}). In this sense, it is similar to the critical value for the 
baryon chemical potential at $T=0$: a first-order transition to nuclear matter occurs 
at $T=0$ when the critical value is reached. We suspect that this class of scenario is not 
likely to be realized in QCD.  For simplicity of language we will refer to the point 
at which the first-order line ends at $T=0$ as a $T=0$ magnetic critical point---even 
though it is not  a critical point in the technical sense---so that we can discuss 
common features for all of the scenarios in a simple way.  We note then that if 
our assumption that the leading-order effective theory has a first-order transition 
at large $B$, a magnetic critical point is guaranteed---either at $T=0$ or finite $T$.
A possible scenario for a phase transition at intermediate $B$ (similar to Fig.~\ref{fig:phase2}) 
was suggested based on a model analysis in Ref.~\cite{Chernodub:2011mc}.

Note that our argument for the magnetic critical point above is applicable not only 
to QCD in the chiral limit, but also to QCD with physical quark masses and QCD with 
any number of flavor with larger quark masses, as long as it has a crossover behavior 
as a function of $T$ at $B=0$.

Lattice QCD studies can distinguish between these scenarios and determine the 
location of the magnetic critical point. As noted earlier, the system does not suffer from 
a fermion sign problem, and practical lattice studies are possible. 
In QCD with realistic quark masses, thermal phase transition was studied
up to $\sqrt{e B} \approx 1 \ {\rm GeV}$ on the lattice \cite{Bali:2011qj}, 
where no signal of the magnetic critical point was found so far; we thus conjecture 
$\sqrt{e B_c} \gtrsim 1 \ {\rm GeV}$ in real QCD. This is not necessarily unusually large, 
remembering the conjectured phase diagram of QCD at finite $\mu_B$, where exotic 
phases (such as color superconducting phases) are expected appear well above the 
nuclear liquid gas phase transition around $\mu_B \approx 1 \ {\rm GeV}$ at $T=0$.
Recall also that the scale of nuclear physics is governed by $\Lambda_{\rm QCD}$ 
plus quark mass corrections, and, e.g., baryonic states in the vacuum have the 
mass comparable to 1 GeV or larger.
For $N_f=2$ QCD with relatively large quark masses ($m_{\pi} \approx 195 \ {\rm MeV}$) 
on the coarse lattice, preliminary evidence for the first-order deconfinement phase 
transition was indicated for $\sqrt{eB} \approx 850 \ {\rm MeV}$ \cite{D'Elia:2010nq}.
This might be consistent with our argument that $B_c$ becomes smaller 
with increasing quark mass, as can be understood from Fig.~\ref{fig:columbia}.

It should be remarked that the phase diagram of QCD in a magnetic field has a 
deep similarity with those of flavor-symmetric QCD at finite isospin chemical 
potential $\mu_I$ \cite{Son:2000xc} and other QCD-like theories at finite $\mu_B$ 
\cite{Kogut:2000ek, Hanada:2011ju}, all of which do \emph{not} have the sign problem; 
quarks acquire a large BCS gap in the superfluid phases of these theories at 
large chemical potential, where no low-lying colored excitation to screen gluons 
exists, and similar deconfined critical points are expected to appear in the 
phase diagrams \cite{Son:2000xc}. However, this is not the case in QCD at finite $\mu_B$. 
This is because the color gauge group is spontaneously broken by a 
color-\emph{nonsinglet} diquark condensate in the color superconductivity at 
sufficiently large $\mu_B$, so that gluons also acquire a mass gap due to the 
Brout-Englert-Higgs mechanism; the low-energy effective theory is {\it not} pure 
gluodynamics in this case. 

The essential conditions for the emergence of the deconfined critical point in 
the $(T, {\cal E})$ plane (with ${\cal E}$ being some parameter of interest, 
such as $B$ and $\mu_I$) are thus a large quark mass gap together with a 
color-\emph{singlet} condensate at large ${\cal E}$. Note that, among these 
theories, the magnetic critical point in the $(T,B)$ plane is the only candidate to 
be relevant, in principle, in nature (as the isospin charge is not conserved due to 
the weak interaction, but the magnetic field does not have such a problem).

Finally, one can ask, from the perspective of theory, whether the magnetic critical point 
continues to exist in the 't Hooft large-$N_c$ limit ($N_c \rightarrow \infty$ with $N_f$ fixed) 
\cite{'tHooft:1973jz}. We here consider QCD with quarks in the fundamental 
representation of the $\SU(N_c)$ gauge group.%
\footnote{The following argument is not applicable to QCD with fundamental quarks 
for fixed $N_f/N_c$ and $N_c \rightarrow \infty$ \cite{Veneziano:1976wm}
and to QCD with adjoint quarks or two-index antisymmetric quarks for fixed 
$N_f$ and $N_c \rightarrow \infty$ \cite{Corrigan:1979xf}, 
as the fermionic degrees of freedom are comparable to the gluonic ones. }
In this limit, the gluon dynamics with $\sim N_c^2$ degrees of freedom is insensitive to 
the quark dynamics with $\sim N_c$. As the magnetic field can only affect the quark dynamics, 
the deconfinement temperature governed by the gluon dynamics is independent of $B$: 
$T_c(B) \sim \Lambda_{\rm QCD}$. (For the $N_f/N_c$ corrections, see Ref.~\cite{Fraga:2012ev}.)
Note here that $\Lambda'_{\rm QCD}(N_c, B) \rightarrow \Lambda_{\rm QCD}$ 
at $N_c \rightarrow \infty$. Therefore, the magnetic critical point does not exist in the 
large-$N_c$ limit. In particular, to study the possible existence of the magnetic critical 
point in the holographic QCD models \cite{Sakai:2004cn}, one needs to incorporate the 
$N_f/N_c$ corrections.

In conclusion, we have argued for a new QCD critical point in the ($T, B$) phase diagram. 
It would be interesting to determine the location of this critical point in lattice 
simulations; as noted above, this should be possible since the theory does not suffer from 
a fermion sign problem.

What is the experimental signature of the finite-$T$ magnetic critical point
in heavy ion collisions (assuming it is  accessible)? This point is characterized by the 
vanishing screening mass of the glueball. Due to the mixing between the glueball $G$ 
and the flavor-singlet meson $\sigma \sim \bar q q$ \cite{Hatta:2003ga}, the singular 
behavior of $G$ is reflected in that of $\sigma$; thus one expects that such 
observables (in the presence of sufficiently small $\mu_B$) are similar to those 
studied for the conventional QCD critical point. Presumably one could distinguish 
between the two by studying the effect as a function of centrality.

More generally, one can consider the phase diagram in the 
three-dimensional space ($T,\mu_B,B$). Whether the magnetic critical point 
persists at large $\mu_B$ would also be an interesting question to be explored.

We thank S.~Aoki, T.~Kanazawa, T.~Kojo, and N.~Su for useful discussions
and F.~Bruckmann, M.~Chernodub, P. de Forcrand, G.~Endrodi, 
M.~Kaminski, and T.~Kovacs for useful comments. 
N. Y. was supported by JSPS Research Fellowships for Young Scientists.


\begin{thebibliography}{99}

\bibitem{Kharzeev:2007jp}
  D.~E.~Kharzeev, L.~D.~McLerran, and H.~J.~Warringa,
  Nucl.\ Phys.\ {\bf A803}, 227 (2008);
  V.~Skokov, A.~Y.~Illarionov and V.~Toneev,
  Int.\ J.\ Mod.\ Phys.\ A {\bf 24}, 5925 (2009);
  W.-T.~Deng and X.-G.~Huang,
  Phys.\ Rev.\ C {\bf 85}, 044907 (2012).

\bibitem{Magnetar} 
R.~C.~Duncan and C.~Thompson, Astrophys. J. {\bf 392}, L9 (1992);
M.~Malheiro, S.~Ray, H.~J.~Mosquera Cuesta, and J.~Dey,
  Int.\ J.\ Mod.\ Phys.\ D {\bf 16}, 489 (2007);
  M.~Eto, K.~Hashimoto, and T.~Hatsuda,
  Phys.\ Rev.\ D {\bf 88}, 081701 (2013).

\bibitem{Grasso:2000wj} 
  D.~Grasso and H.~R.~Rubinstein,
  Phys.\ Rept.\  {\bf 348}, 163 (2001).

\bibitem{Bali:2011qj} 
  G.~S.~Bali, F.~Bruckmann, G.~Endrodi, Z.~Fodor, S.~D.~Katz, S.~Krieg, A.~Schafer, and K.~K.~Szabo,
  JHEP {\bf 1202}, 044 (2012);
  Phys.\ Rev.\ D {\bf 86}, 071502 (2012).

\bibitem{Mizher:2010zb} 
  A.~J.~Mizher, M.~N.~Chernodub, and E.~S.~Fraga,
  Phys.\ Rev.\ D {\bf 82}, 105016 (2010);
  R.~Gatto and M.~Ruggieri,
  Phys.\ Rev.\ D {\bf 82}, 054027 (2010);
  Phys.\ Rev.\ D {\bf 83}, 034016 (2011);
  K.~Kashiwa,
  Phys.\ Rev.\ D {\bf 83}, 117901 (2011).

\bibitem{Fukushima:2012kc} 
  K.~Fukushima and Y.~Hidaka,
  Phys.\ Rev.\ Lett.\  {\bf 110}, 031601 (2013);
  J.~O.~Andersen and A.~A.~Cruz,
  Phys.\ Rev.\ D {\bf 88}, 025016 (2013);
  T.~Kojo and N.~Su,
  Phys.\ Lett.\ B {\bf 720}, 192 (2013);
  F.~Bruckmann, G.~Endrodi, and T.~G.~Kovacs,
  JHEP {\bf 1304}, 112 (2013).

\bibitem{Galilo:2011nh} 
  B.~V.~Galilo and S.~N.~Nedelko,
  Phys.\ Rev.\ D {\bf 84}, 094017 (2011);
  E.~S.~Fraga and L.~F.~Palhares,
  Phys.\ Rev.\ D {\bf 86}, 016008 (2012).

\bibitem{Fraga:2012ev} 
  E.~S.~Fraga, J.~Noronha, and L.~F.~Palhares,
  Phys.\ Rev.\ D {\bf 87}, 114014 (2013).

\bibitem{Anber:2013tra} 
  M.~M.~Anber and M.~Unsal,
  arXiv:1309.4394 [hep-th].

\bibitem{Agasian:2008tb} 
  N.~O.~Agasian and S.~M.~Fedorov,
  Phys.\ Lett.\ B {\bf 663}, 445 (2008).

\bibitem{Stephanov:2004wx} 
  M.~A.~Stephanov,
  Prog.\ Theor.\ Phys.\ Suppl.\  {\bf 153}, 139 (2004)
  [Int.\ J.\ Mod.\ Phys.\ A {\bf 20}, 4387 (2005)].

\bibitem{Hatsuda:2006ps} 
  T.~Hatsuda, M.~Tachibana, N.~Yamamoto, and G.~Baym,
  Phys.\ Rev.\ Lett.\  {\bf 97}, 122001 (2006);
  Phys.\ Rev.\ D {\bf 76}, 074001 (2007).

\bibitem{Andersen:2012bq} 
  J.~O.~Andersen and A.~Tranberg,
  JHEP {\bf 1208}, 002 (2012);
  M.~Ruggieri, M.~Tachibana, and V.~Greco,
  arXiv:1305.0137 [hep-ph].




\bibitem{Shovkovy:2012zn} 
  I.~A.~Shovkovy,
  Lect.\ Notes Phys.\  {\bf 871}, 13 (2013).

\bibitem{Miransky:2002rp} 
  V.~A.~Miransky and I.~A.~Shovkovy,
  Phys.\ Rev.\ D {\bf 66}, 045006 (2002).


\bibitem{Yaffe:1982qf}
  L.~G.~Yaffe and B.~Svetitsky,
  Phys.\ Rev.\ D {\bf 26}, 963 (1982);
  B.~ Svetitsky, Phys.~Rept. {\bf 132}, 1 (1986).

\bibitem{Fukugita:1989yb} 
  M.~Fukugita, M.~Okawa, and A.~Ukawa,
  Phys.\ Rev.\ Lett.\  {\bf 63}, 1768 (1989);
  Nucl.\ Phys.\ {\bf B337}, 181 (1990).

\bibitem{Son:1998uk} 
  D.~T.~Son,
  Phys.\ Rev.\ D {\bf 59}, 094019 (1999).

\bibitem{Rischke:2000cn} 
  D.~H.~Rischke, D.~T.~Son, and M.~A.~Stephanov,
  Phys.\ Rev.\ Lett.\  {\bf 87}, 062001 (2001).

\bibitem{Aoki:2006we} 
  Y.~Aoki, G.~Endrodi, Z.~Fodor, S.~D.~Katz, and K.~K.~Szabo,
  Nature {\bf 443}, 675 (2006).


\bibitem{Chernodub:2011mc} 
  M.~N.~Chernodub,
  Phys.\ Rev.\ Lett.\  {\bf 106}, 142003 (2011).

\bibitem{D'Elia:2010nq} 
  M.~D'Elia, S.~Mukherjee, and F.~Sanfilippo,
  Phys.\ Rev.\ D {\bf 82}, 051501 (2010).

\bibitem{Son:2000xc}
  D.~T.~Son and M.~A.~Stephanov,
  Phys.\ Rev.\ Lett.\  {\bf 86}, 592 (2001).

\bibitem{Kogut:2000ek} 
  J.~B.~Kogut, M.~A.~Stephanov, D.~Toublan, J.~J.~M.~Verbaarschot, and A.~Zhitnitsky,
  Nucl.\ Phys.\ {\bf B582}, 477 (2000).

\bibitem{Hanada:2011ju} 
  M.~Hanada and N.~Yamamoto,
  JHEP {\bf 1202}, 138 (2012).

\bibitem{'tHooft:1973jz}
  G.~'t Hooft,
  Nucl.\ Phys.\ {\bf B72}, 461 (1974).

\bibitem{Veneziano:1976wm}
  G.~Veneziano,
  Nucl.\ Phys.\  {\bf B117}, 519 (1976).
  
\bibitem{Corrigan:1979xf}
  E.~Corrigan and P.~Ramond,
  Phys.\ Lett.\  {\bf B87}, 73 (1979).

\bibitem{Sakai:2004cn} 
  T.~Sakai and S.~Sugimoto,
  Prog.\ Theor.\ Phys.\  {\bf 113}, 843 (2005).

\bibitem{Hatta:2003ga} 
  Y.~Hatta and K.~Fukushima,
  Phys.\ Rev.\ D {\bf 69}, 097502 (2004).

\end{thebibliography}
\end{document}